\documentclass[prd,aps,twocolumn,nofootinbib,longbibliography]{revtex4}
\usepackage{graphicx}
\usepackage{hyperref}
\usepackage{amsmath,amssymb}
\usepackage{xcolor}

\definecolor{darkBlue}{rgb}{0, 0, 0.8}

\hypersetup{
  bookmarksopen=true,     
  colorlinks=true,        
  linkcolor=darkBlue,     
  citecolor=darkBlue,      
  filecolor=darkBlue,      
  urlcolor=darkBlue        
}

\def\be{\begin{equation}}
\def\ee{\end{equation}}
\def\ba{\begin{align}}
\def\ea{\end{align}}

\def\gr{$\gamma$-ray}

\def\lsim{\raise0.3ex\hbox{$\;<$\kern-0.75em\raise-1.1ex\hbox{$\sim\;$}}}
\def\gsim{\raise0.3ex\hbox{$\;>$\kern-0.75em\raise-1.1ex\hbox{$\sim\;$}}}

\def\theta{\vartheta}

\definecolor{purple}{rgb}{0.5,0,0.5}

\newcommand{\change}[1]{\textcolor{black}{#1}}

\begin{document}

\title{Energy spectra of secondaries in proton-proton interactions}

\author{S.~Koldobskiy$^{1,2}$}
\author{M.~Kachelrie\ss$^3$}
\author{A.~Lskavyan$^{1}$}
\author{A.~Neronov$^{4,5}$}
\author{S.~Ostapchenko$^{6,7}$}
\author{D.~V.~Semikoz$^{1,4,8}$}

\affiliation{$^{1}$National Research Nuclear University MEPHI, 115409 Moscow, Russia}
\affiliation{$^{2}$Space Physics and Astronomy Research Unit and Sodankylä Geophysical Observatory, University of Oulu, 90014 Oulu, Finland}
\affiliation{$^{3}$Institutt for fysikk, NTNU, Trondheim, Norway}
\affiliation{$^{4}$APC, Universit\'e Paris Diderot, CNRS/IN2P3, CEA/IRFU, Observatoire de Paris, Sorbonne Paris Cit\'e, 119 75205 Paris, France}
\affiliation{$^{5}$Astronomy Department, University of Geneva, Ch. d'Ecogia 16, 1290, Versoix, Switzerland}
\affiliation{$^{6}$II.\ Institute for Theoretical Physics, Hamburg University, Hamburg, Germany}
\affiliation{$^{7}$D.V. Skobeltsyn Institute of Nuclear Physics, Moscow State University, Moscow, Russia}
\affiliation{$^{8}$INR RAS, 60th October Anniversary prospect 7a, Moscow, Russia}

\begin{abstract}
We compare the predictions of {\tt AAfrag} for the spectra of secondary
photons, neutrinos, electrons, and positrons produced in proton-proton
collisions to those of the parameterisations of  Kamae {\it et al.}, Kelner
{\it et al.} and Kafexhiu {\it et al.\/}  We find that the differences in
the normalisation of the photon energy spectra reach  20--50\%  at
intermediate values of the  transferred energy fraction $x$, growing up to a
factor of two for $x\to 1$, while the differences in the neutrino spectra
are even larger. We argue that LHCf results on the forward production of
photons \change{and neutral pions} favor the use of the QGSJET-II-04m model on which {\tt AAfrag} is
based. The differences in the normalisation have important implications
in the  context of multi-messenger astronomy, in particular, for the
prediction of neutrino fluxes, based on gamma-ray flux measurements, or
regarding the inference of the cosmic ray spectrum, based on gamma-ray data.
We note also that the positron-electron ratio from hadronic interactions
increases with energy towards the cutoff, an effect which is missed using
the  average electron-positron spectrum from Kelner~{\it et al.} 
Finally, we describe the publicly available python package {\tt aafragpy},
which provides the  secondary spectra of photons, neutrinos, electrons, and
positrons. This package  complements the  {\tt AAfrag} results for protons
with energies above 4\,GeV with previous analytical parameterisations of
particle spectra for lower energy protons. 
\end{abstract}

\date{\today}

\maketitle

\section{Introduction}

A large variety of applications in astroparticle physics relies on the precise
knowledge of the production cross sections of secondary particles in hadronic
interactions. A prominent example is the branch of multi-messenger
astronomy which aims to connect photon, neutrino and cosmic ray
data~\cite{Spurio:2018knn,Kachelriess:2019oqu}.
While in the past simple estimates or empirical parameterisations for these
cross sections were sufficient, the improved accuracy and large statistics
of current and future experiments like AMS-02~\cite{Aguilar:2021tos},
LHASSO~\cite{Bai:2019khm}, CTA~\cite{CTAConsortium:2018tzg} and
IceCube-Gen2~\cite{Aartsen:2020fgd}
requires a corresponding advancement of the theoretical predictions.

There exist two main approaches to the description of hadronic production
cross sections. In the first one, one parametrises hadronic interaction data
using empirical scaling laws. 
In spite of its convenience, the use of such parameterisations
becomes dangerous when the latter are extrapolated outside the kinematical 
range of the data, they are based on.
In particular, high-energy extrapolations into
the multi-PeV range of such empirical parameterizations are generally
unreliable, but are required for the correct interpretation of gamma-ray and
neutrino data in the 100\,TeV--1\,PeV energy range.
As an alternative, one can use QCD inspired Monte Carlo event generators for
the description of hadronic interactions. In order to provide a fast and
user-friendly tool for the computation of the production cross sections, one
can first bin their results and then either  interpolate or fit them. Given
sufficient statistics, the former option reproduces exactly the results of the
used Monte Carlo event generator, while the accuracy of the latter approach
depends on the choice of appropriate fit functions. Similar to the case of
empirical parameterisations, the use of such fit functions becomes dangerous
when they are extrapolated outside the fit range.

While most of the QCD inspired Monte Carlo event generators used in cosmic ray
physics were overall in a satisfactory agreement~\cite{dEnterria:2011twh} with
various data from Run~I of the Large Hadron Collider (LHC), several of
them have been updated by re-tuning their model parameters with LHC data.
Moreover, the QGSJET-II-04 model~\cite{Ostapchenko:2010vb,Ostapchenko:2013pia}
was further tuned in Ref.~\cite{Kachelriess:2015wpa}
to improve antiproton production at low energies. The results of this
QGSJET-II-04m tune were used to provide 
{\tt AAfrag}~\cite{Kachelriess:2019ifk,AAfrag} with convenient
tabulations of the production cross-sections of secondary particles 
in proton-proton, proton-nucleus, nucleus-proton, and nucleus-nucleus
reactions.

The aim of this work is to compare the predictions of various available
parameterisations for the spectra of secondary photons, neutrinos, electrons,
and positrons produced in proton-proton collisions to those of {\tt AAfrag}.
The predictions of these parameterisations are expected to vary mainly
because of differences in the physics of the event generators used
to produce the fit data. Moreover, parametrisations based on pre-LHC
and post-LHC event generators are expected to differ significantly,
especially for forward
particle production  and in the multi-PeV energy range. In addition, the
treatment of the experimental data and the choice of the fit
functions affects the predicted spectra. As a result of these differences,
we find significant variations between the results of these
parametrisations and we characterise the observed discrepancies.
In addition, we describe and make publicly available  the python
package {\tt AAfragpy}, which provides the  secondary spectra of
\gr s, neutrinos, electrons and positrons for primary proton energies
in the energy  range 1.5--$10^{11}$\,GeV. Since {\tt AAfrag} is restricted to
proton energies above 4\,GeV, we complement it in this python package at
lower energies with the parameterisations of
Kamae {\it et al.\/}~\cite{Kamae:2006bf}.

This article is organised as follows: We start by describing in
Section~\ref{sec:models}
the main features of the parametrisations we  examine. In
Section~\ref{sec:crosssection}, we compare secondary particle production
for fixed energies of the incident  protons, while we discuss the case of
an $1/E^2$ primary proton  spectrum in Section~\ref{sec:spectra_gam_nu}. After
a presentation of the new python package {\tt aafragpy} in
Section~\ref{sec:py}, we conclude.

\section{Parametrisations}
\label{sec:models}

In addition to {\tt AAfrag}, we discuss the three  most commonly\footnote{For a discussion of other models see Ref.~\cite{Kafexhiu:2014cua}}
used parametrisations of~\citet{Kamae:2006bf},
\citet{Kelner:2006tc}, and
\citet{Kafexhiu:2014cua}. Their main characteristics 
are summarized in Table~\ref{tab:models}. A drawback of QCD
inspired event generators is that they cannot be used below
a minimal energy, which is typically in the range of few to 100\,GeV for the
energy $E_p$ of the projectile in the lab frame. Note that the recommended minimal energy
is with 56 and 100\,GeV for Pythia and SIBYLL, respectively, rather high.
In contrast, it was shown in Ref.~\cite{Kachelriess:2019ifk}
that QGSJET-II-04m can be used down to primary energies as low as 4\,GeV.
All  three parametrisations complement  fit functions from
event generators at high energies with phenomenological models at low
energies.  Kamae {\it et al.\/} used several parameterized models, including
resonance-excitation components for primary energies $E_p$ below 52.6\,GeV.
Kelner {\it et al.\/} proposed to use a $\delta$ function approximation for
the calculation of the production spectra of secondaries
at $E_p\leq 100$\,GeV or for $E_{\rm s}/E_p\leq10^{-3}$,  where $E_{\rm s}$
denotes the energy of the secondary particle of interest. This approximation
is restricted to the case of a power-law spectra in momentum of the proton
primaries and the (invalid) assumption of a constant inelastic cross section.
Kafexhui {\it et al.\/} relied on a
compilation of  experimental data below $E_p<2$\,GeV and applied GEANT~4.10.0
at intermediate energies.
These latter authors considered only the production of photons.
Moreover, they re-used the results for the meson spectra from
Kelner {\it et al.\/} in the case of QGSJET-I and SIBYLL~2.1, but applied
different fit functions for the resulting photon spectra. Thus one can
view the differences between the results of Kafexhui {\it et al.\/} and
Kelner {\it et al.\/} as a measure for the deviations introduced by the
fit procedure. Note also that
SIBYLL~2.1 was released in the year 2000, while the version of QGSJET-I
used in Refs.~\cite{Kelner:2006tc,Kafexhiu:2014cua} was published in 1997.
Similarly, Pythia~6.2 is a pre-LHC event generator dating from the year 2001.

\begin{table}
\caption{Characteristics of the models.}
\begin{tabular}{ c|c|c} 
	\hline
	\hline
	        & High-energy   & Range in kin. energy, \\
		&  interaction model	 &  GeV  \\
	\hline
	This work 								& QGSJET-II-04m & 3.1\,--\,$10^{11}$\\ 
	Kamae {\it et al.\/}  		& Pythia 6.2 & 0.488\,--\,$5.12\times 10^5$ \\
        Kelner {\it et al.\/}  	& SIBYLL~2.1 & 100\,--\,10$^8$ \\ 
	Kafexhiu {\it et al.\/}  & GEANT 4.10.0 & 0.280\,--\,10$^5$\\ 
													  & PYTHIA 8.18& 50\,--\,10$^6$ \\ 
													 & QGSJET-I& 100\,--\,10$^6$  \\ 
													 & SIBYLL 2.1 & 100\,--\,10$^6$ \\ 
	\hline
	\hline
\end{tabular}
\label{tab:models}
\end{table}

\section{Energy spectra  for monoenergetic protons}
\label{sec:crosssection}

\subsection{Photons and neutrinos}

\begin{figure}
\includegraphics[width=\linewidth,angle=0]{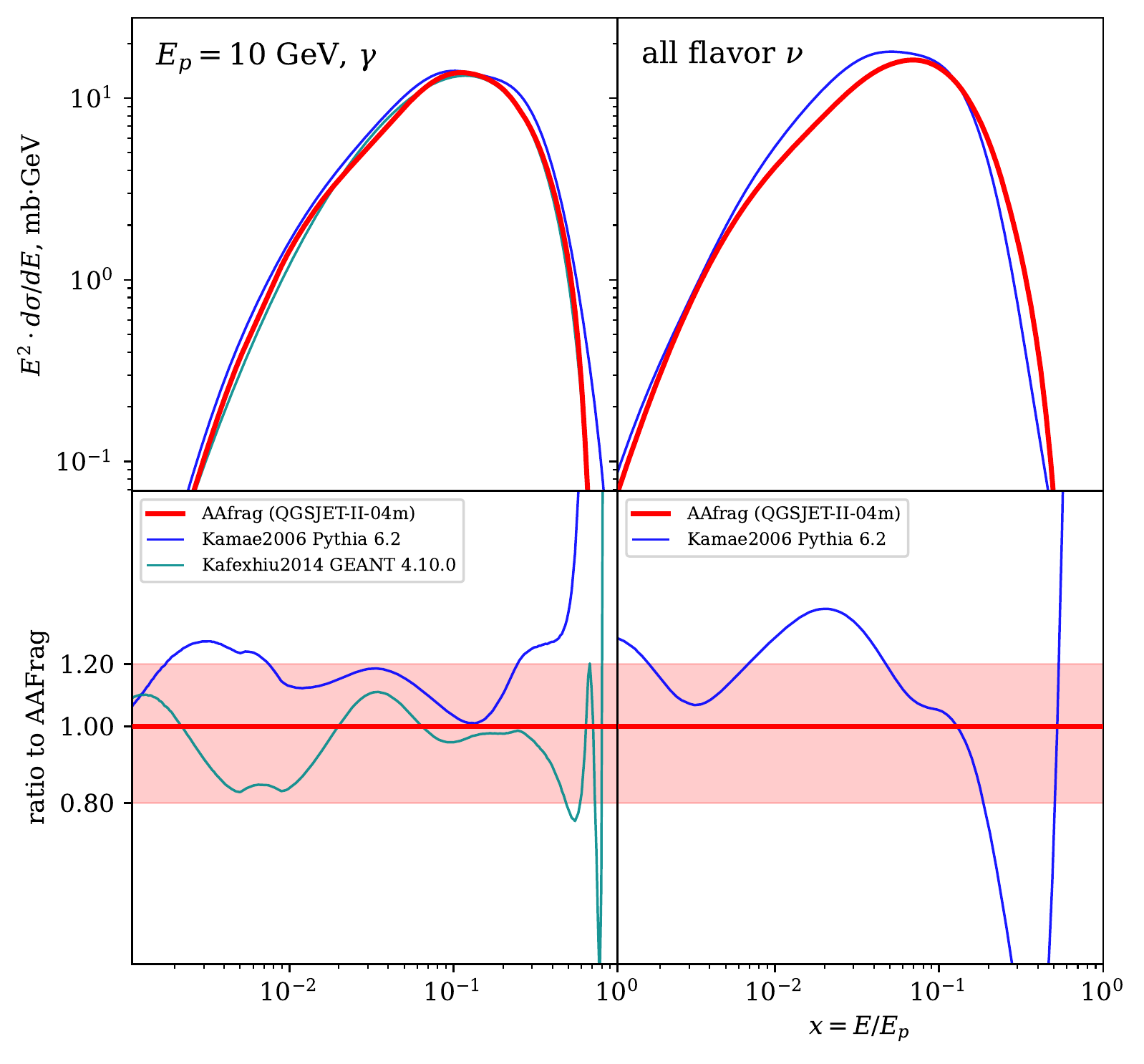}
\caption{Gamma-ray (left) and neutrino (right) spectra
  $E^2{\rm d}\sigma /{\rm d}E$ as a function of the transferred energy
  fraction $x=E/E_p$, for 10\,GeV incident  protons. The upper left panel
  compares the photon spectra calculated with {\tt AAfrag} with those of
  Kamae~{\it et al.} based on PYTHIA~6.2 and Kafexhiu~{\it et al.} based on
  GEANT~4.10.0. The spectra of all-flavour neutrinos 
  are compared to Kelner~{\it et al.} and Kamae~{\it et al.}
  The lower panels show the ratio of the spectra to the
  {\tt AAfrag} results.
  \label{fig:gamma10GeV}}
\end{figure}

\begin{figure}
\includegraphics[width=\linewidth,angle=0]{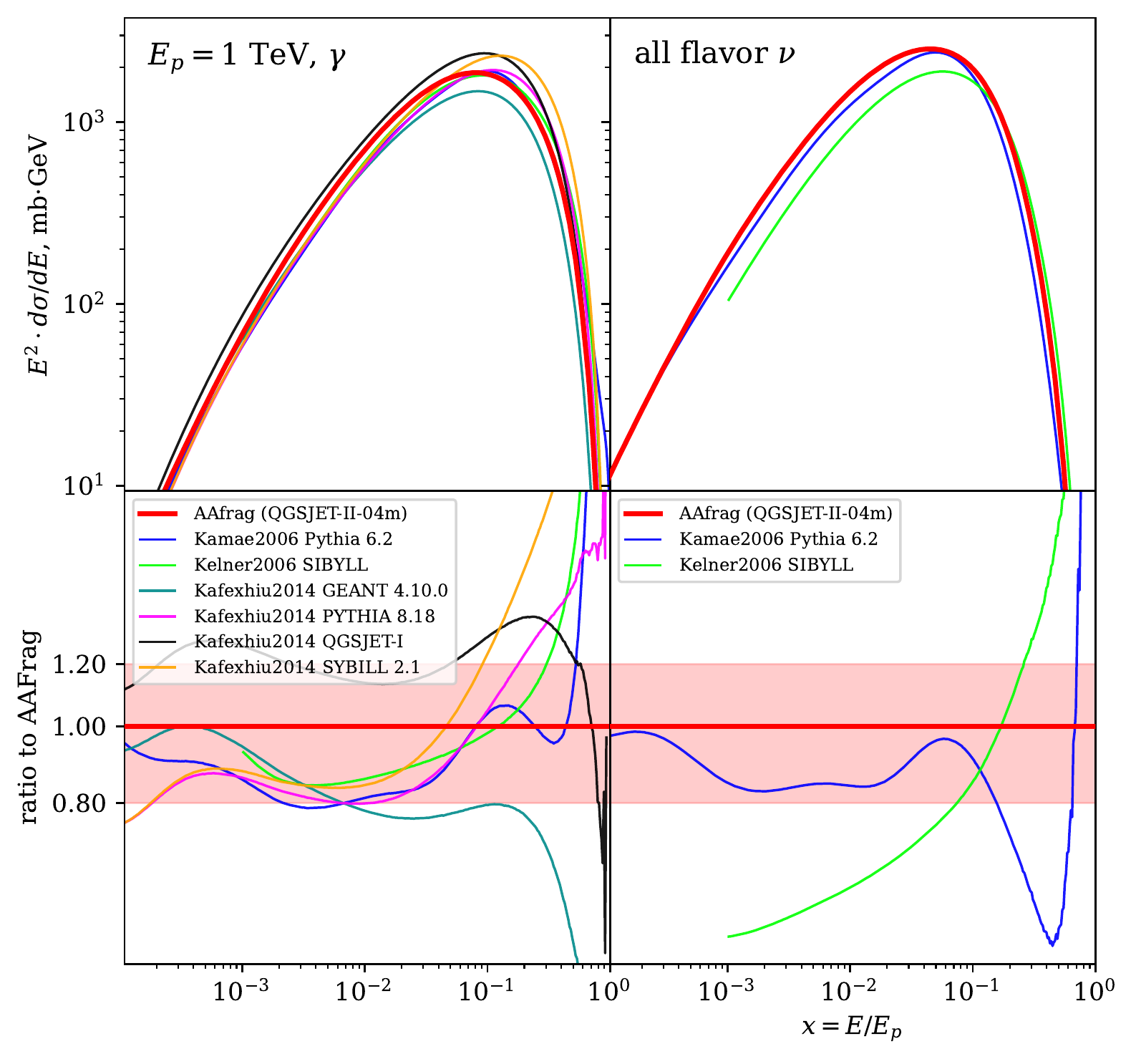}
\caption{Gamma-ray and neutrino spectra $E^2{\rm d}\sigma /{\rm d}E$ (top)
  and ratios (bottom) as a function of the transferred energy
  fraction $x=E/E_p$, for 1\,TeV incident  protons. The parametrisations
  of Kelner~{\it et al.\/}  based on SIBYLL~2.1 and Kafexhiu~{\it et al.\/}
  based on SIBYLL, PYTHIA and QGSJET-I are added to the analysis.
\label{fig:gammaTeV}}
\end{figure}

\begin{figure}
\includegraphics[width=\linewidth,angle=0]{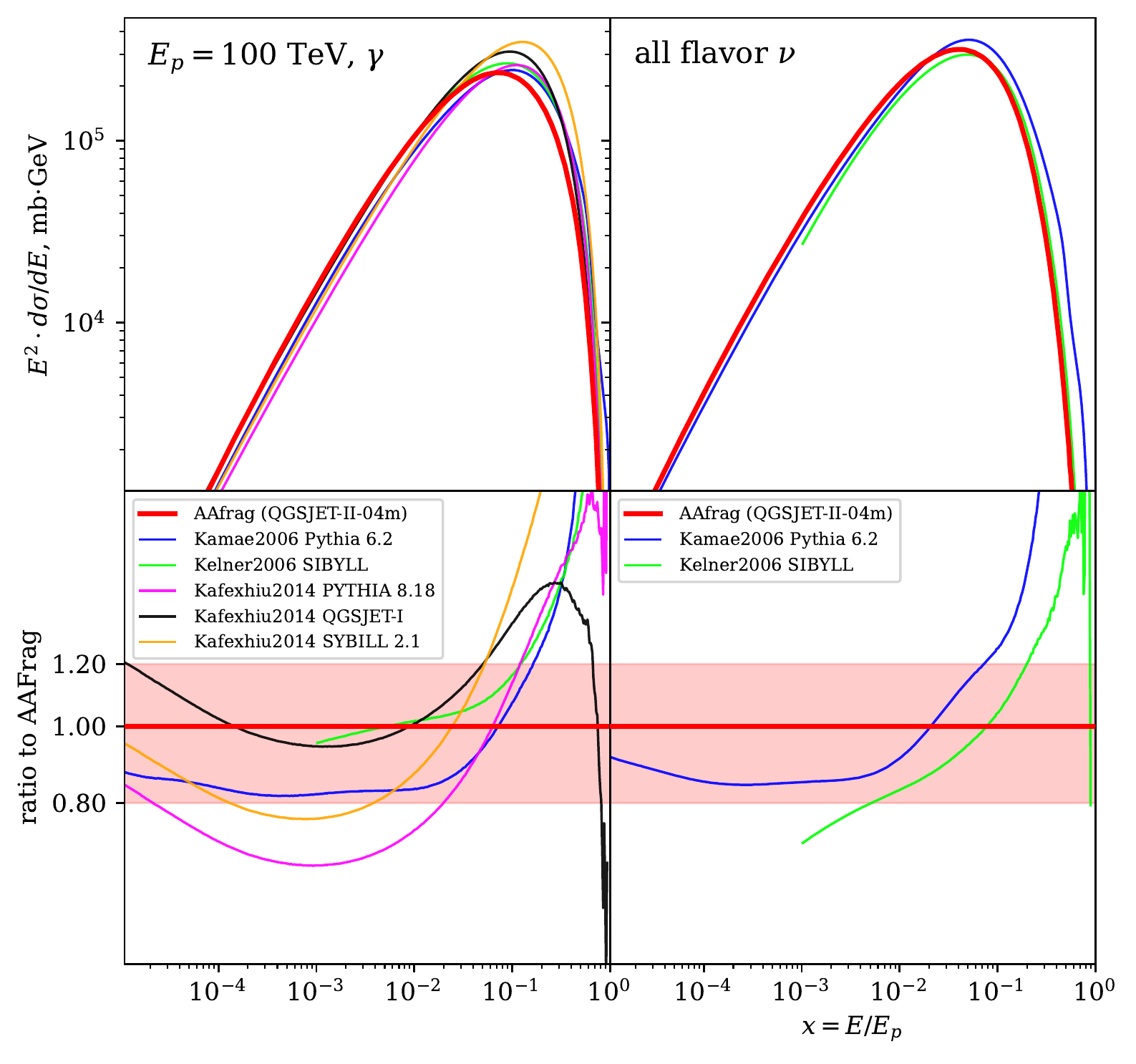}
\caption{Same as Fig.~\ref{fig:gammaTeV}  for 100\,TeV primary 
protons.
\label{fig:gamma100TeV}}
\end{figure}

We start our comparison with the spectra $E^2 {\rm d}\sigma /{\rm d}E$ of
photons and neutrinos, produced in proton-proton interactions with fixed primary
energies. These spectra, which correspond to the spectral energy distribution
(SED) of hypothetical sources of monoenergetic protons, are shown in 
Figs.~\ref{fig:gamma10GeV}, \ref{fig:gammaTeV}, and \ref{fig:gamma100TeV}.
In Fig.~\ref{fig:gamma10GeV}, which
corresponds to the incident proton energy  $E_p=10$\,GeV, we  observe a 20\% 
scatter in the model results for the photon spectra at the peak of the SED,
in the photon energy range $10^{-2}E_p<E_\gamma<0.1E_p$. 
 The differences between the model predictions grow up to $\sim 50\%$ in the 
high-energy tails of the SED, which reflects the lack of experimental data for
very forward photon production at these energies.  
The calculated  neutrino production spectra shown
in the right panel of Fig.~\ref{fig:gamma10GeV} deviate somewhat stronger,
with $\sim 30$\% differences. In particular, the SED of {\tt AAfrag} is more
sharply peaked in the forward direction than the one of Kamae {\it et al.\/},
which reflects the harder spectra of charged pions in QGSJET-II-04m
compared to \change{that parametrization.}

\begin{figure}
\includegraphics[width=0.9\linewidth,angle=0]{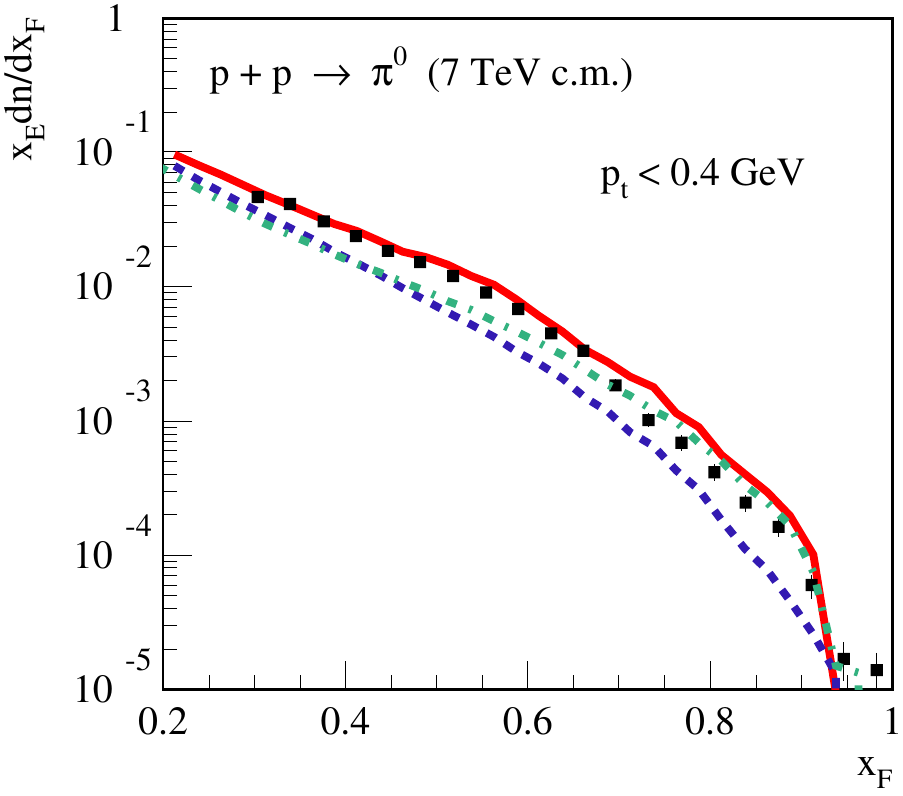}
\caption{Measurements of the  differential cross-section of $\pi^0$ production
  at 7\,TeV c.m.\ energy as function of Feynman $x_F$  by LHCf (black
  squares with errorbars) compared to the predictions of QGSJET-II-04m
  (solid red), SIBYLL 2.1 (dashed-dotted green) and QGSJET-I (dashed blue). 
\label{fig:LHCf}}
\end{figure}

Moving to higher energies, the differences  between {\tt AAfrag} and earlier
models stay within 20\% at the peak of the SED of secondary particles, cf.\
with Figs.~\ref{fig:gammaTeV} and \ref{fig:gamma100TeV}. On the other hand,
the results of the different calculations deviate much stronger in the forward
direction, for  $x=E_s/E_p \gtrsim 0.1$, where differences exceed 50\% for
some parametrisations. For
the particular case of the model of  Kamae {\it et al.\/}, this deviation
is  related to deficiencies in their modelling of inelastic diffraction,
as discussed in some detail in Ref.~\cite{Kachelriess:2012fz}; the
corresponding contribution is visible by eye as the sharp rise of the
blue lines for  $E_{\gamma} \rightarrow E_p$
in the lower left panels of Figs.~\ref{fig:gammaTeV} and \ref{fig:gamma100TeV}.
On the other hand, the softer photon spectra for $E_{\gamma} \rightarrow E_p$
obtained in the  parametrization of  Kafexhui {\it et al.\/}, based on the
outdated QGSJET-I model~\cite{Kalmykov:1993qe}, are caused by the too soft
pion production spectra in that model. Overall, the increase with
energy of the differences between the various predictions for the forward
production of secondary particles is due to the fact that relevant experimental
data have till recently been available at fixed target energies only, for 
$E_p\leq 400$ GeV. This lack of experimental data has been especially
unsatisfactory, because the role of forward production is greatly enhanced
in the case of a steep spectrum of primary cosmic rays, as discussed in some
detail in  Refs.~\cite{Kachelriess:2012fz,Kachelriess:2014mga}; this issue
will be in this work further addressed in Sec.~\ref{sec:spectra_gam_nu} below.
Therefore, the measurements of forward photon and neutral pion production at
LHC  energies, $\sqrt{s}=0.9$, 2.76, 7, and 13\,TeV, by the LHCf
experiment~\cite{Adriani:2011nf,Adriani:2012fz,Adriani:2015iwv,Adriani:2017jys}
are of great importance. 
\change{In Fig.~\ref{fig:LHCf}, we compare  LHCf measurements (black
  squares with errorbars) of the differential cross-section of $\pi^0$
  production at 7\,TeV c.m.\ energy as function of Feynman $x_F$  to the
  predictions of QGSJET-II-04m  (solid red line), SIBYLL 2.1 (dashed-dotted green)
  and QGSJET-I (dashed blue). It can be seen that  QGSJET-II-04m describes
  overall the data best. The only parametrisation which provides enough
  kinematical information that it can be compared to such measurements
  is the one of Kamae {\it et al.}  While 7\,TeV c.m.\ energy is 
  beyond the range of applicability of this parametrisation, the predicted
  yield of photons at 900\,GeV c.m.\ energy  is already a factor few above
  the LHCf measurements. 
  Therefore, 
  the parametrizations of
  Kamae~{\it et al.\/} and Kafexhui~{\it et al.\/} 
  (blue, green,  and magenta lines in Figs.~\ref{fig:gammaTeV} and
  \ref{fig:gamma100TeV}) are disfavored by the LHCf data.} Note also that the
difference between the results of Kafexhui~{\it et al.\/} 
and Kelner~{\it et al.\/}, both based on SIBYLL 2.1,
 which is caused purely by the use of different fit
functions, is as large as 20\% at $x\simeq 0.1$. Moreover, we observe
a rather large discrepancy with the Kelner {\it et al.\/} parametrisation
for the neutrino spectra, where  both the normalisation and the shape of
the neutrino spectrum  differ significantly from the other predictions.
As a result, we expect correspondingly large errors in the predictions of
the expected number of observable neutrino events from specific sources,
which are  based on the Kelner~{\it et al.\/} parametrisation.

\begin{figure}
\includegraphics[width=\linewidth,angle=0]{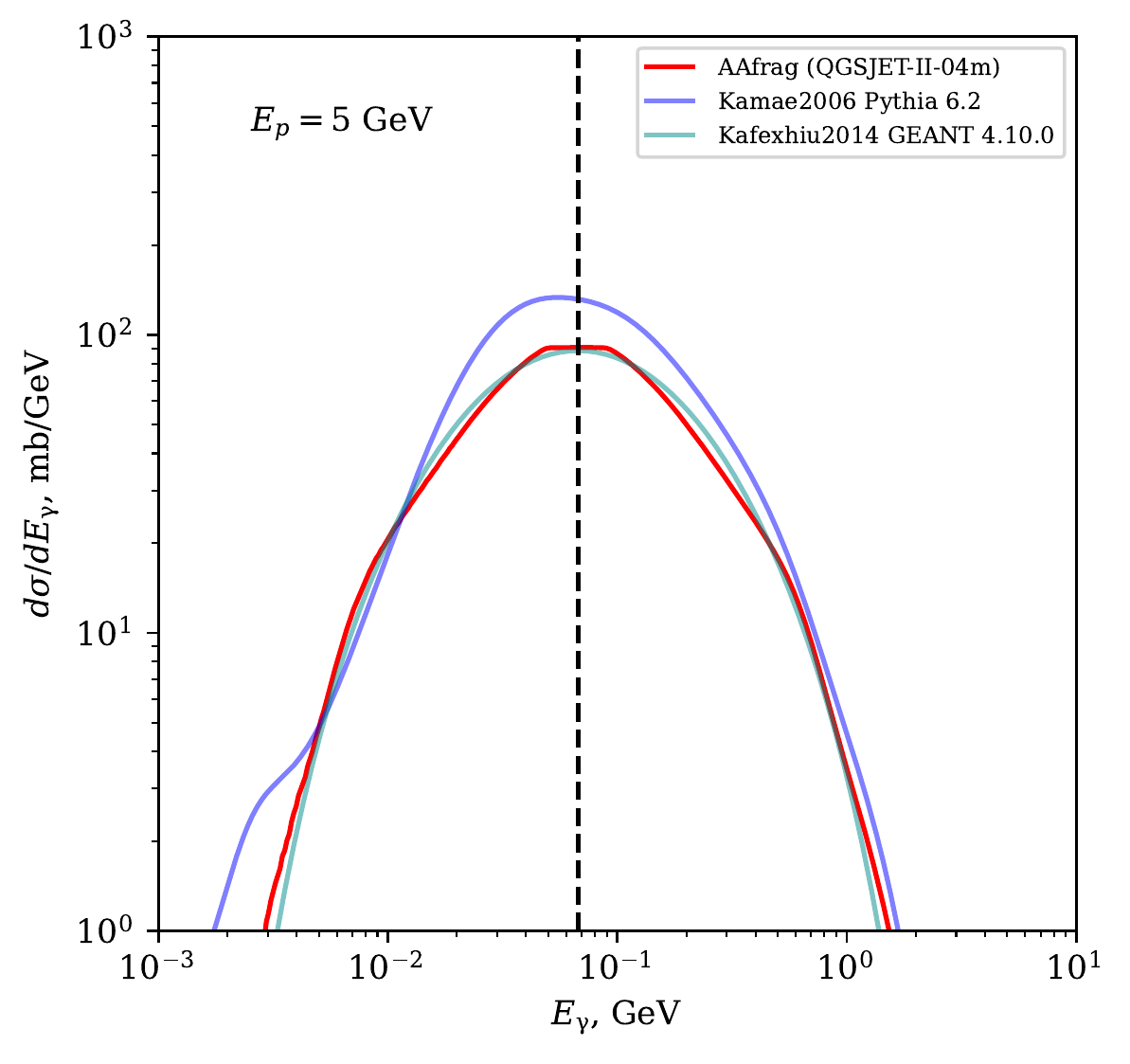}
\caption{Differential cross-section of $\gamma$-ray production for 5\,GeV protons calculated  with {\tt AAfrag}, Kamae~{\it et al.} and Kafexhiu~{\it et al.} codes.
\label{fig:gamma5}}
\end{figure}

Figures~\ref{fig:gamma10GeV}, \ref{fig:gammaTeV}, and
\ref{fig:gamma100TeV}  also show that the predicted spectral shapes of the photon SED 
in the limit of small $E_{\gamma}$ differ  for the various parametrizations. 
To clarify this point, we plot in Fig.~~\ref{fig:gamma5} the photon production
spectrum ${\rm d}\sigma /{\rm d}E$ as function of $\log(E)$ for $E_p=5$\,GeV,
calculated using both {\tt AAfrag} and the parametrizations. Plotted in such
a way, the photon spectrum should be symmetric with respect to the energy
$E_\gamma=m_{\pi^0}/2\simeq 67.5$\,MeV~\cite{Stecker71}. While this is the case
for the Kafexhiu~{\it et al.\/} results based on GEANT~4.10.0 and for
{\tt AAfrag}, this symmetry is broken  by the employed
fit functions in the case of the parametrizations.
It is worth recalling, however, that the high energy gamma-ray fluxes from
astrophysical sources are dominated by the forward photon production in
cosmic ray interactions (see, e.g.\ Ref.~\cite{Kachelriess:2014mga}).
Therefore,  the low energy part of the photon production spectra is only 
relevant to the calculation of photon spectra in the sub-GeV energy range,
where the important contributions are coming from proton interactions at
relatively low energies.

\subsection{Electrons and positrons}

\begin{figure}
\includegraphics[width=1.04\linewidth,angle=0]{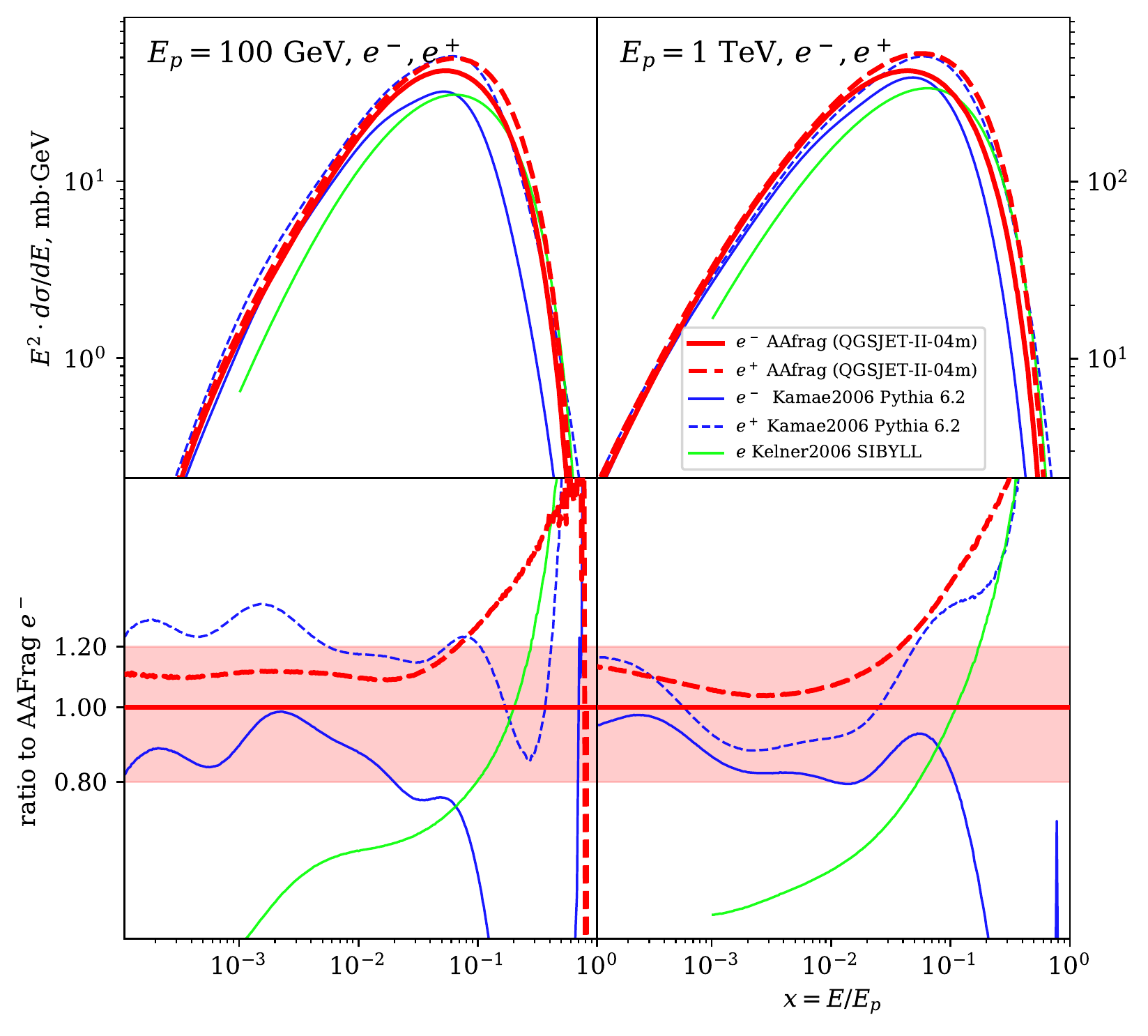}
\caption{Electron and positron spectra
  $E^2{\rm d}\sigma /{\rm d}E$ as a function of the transferred energy
  fraction $x=E/E_p$, for 100\,GeV and 1\,TeV incident  protons.
\label{fig:elpos_1TeV}}
\end{figure}

In addition to photons and neutrinos, we plot in Fig.~\ref{fig:elpos_1TeV} the production
spectra of positrons and electrons for $E_p=$\,100 GeV and 1 TeV, 
calculated using both {\tt AAfrag} and the parametrizations.
Here, apart from the general differences between the various parametrizations,
it is important to note that, unlike   {\tt AAfrag} and  Kamae {\it et al.\/},
the parametrization of Kelner {\it et al.\/}
 provides average spectra between $e^+$ and $e^-$.
That way, one neglects the important difference between them, which stems 
from significantly harder production spectra of $\pi^+$ mesons, compared to  
$\pi^-$, in proton-proton interactions. As will be further discussed in
Section~\ref{sec:electron_spectra}, this has important consequences for the
analysis and the interpretation of experimental data on cosmic-ray fluxes of electrons
and positrons.

\section{Energy spectra from broad energy distributions of protons}
\label{sec:spectra_gam_nu}

The intensity $I_a$ of secondary particles of type $a$ is related to the
intensity of primary cosmic rays $I_p$ as
\begin{equation}
  I_a(E) =\int_0^\infty {\rm d}l\, n_{\rm gas} \int_{E}^{\infty} {\rm d}E'\,
  \frac{{\rm d}\sigma_a}{{\rm d}E}(E',E) I_p(E')  ,
\end{equation}
where the $l$ integration is along a fixed line-of-sight and $n_{\rm gas}$
denotes the number density of target protons. The cosmic-ray
spectrum $I_p$ is expected to follow in most of  sources roughly a power-law
in momentum, modified by a high-energy cutoff which is often chosen as an
exponential,
\begin{equation}
  I_p(p) = p^{-\gamma} \exp (-p/p_0).
\end{equation}
In the case of diffusive acceleration on non-relativistic supersonic shocks,
one expects $\gamma\simeq 2.0$--2.2~\cite{Bell:2013vxa}. In the following examples, we will use
$\gamma=2$ and use as energy of the cut-off $p_0=100$\,TeV. Additionally,
we will impose a  sharp low-energy cut-off at $E_p=4$\,GeV, since  the
{\tt AAfrag} predictions are available only above this  energy.

\subsection{Photon and neutrino spectra}

\begin{figure}
\includegraphics[width=\linewidth]{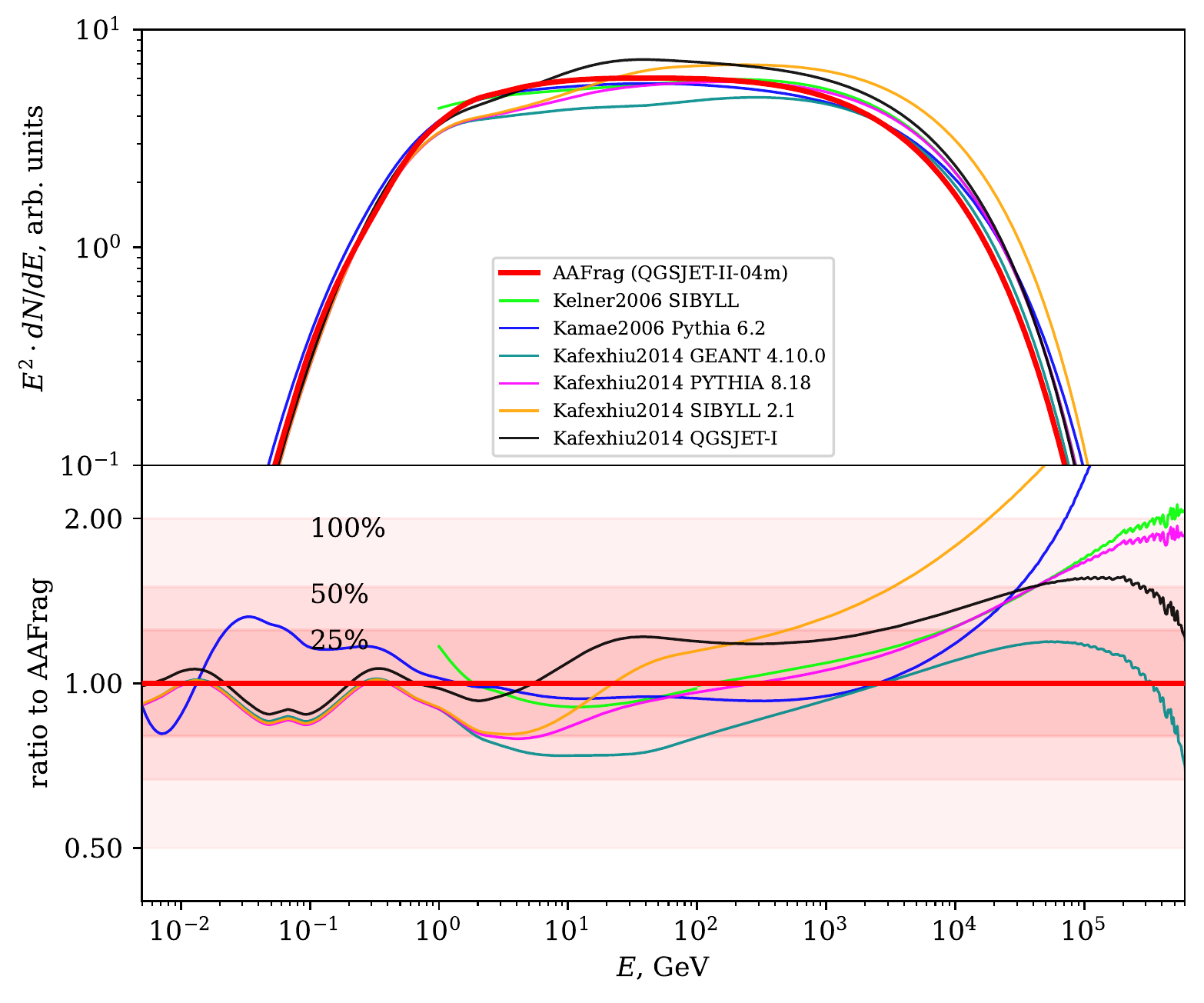}
\caption{Gamma-ray energy flux for a power law primary proton spectrum, with an 
exponential cutoff,  $\propto 1/p^2 \exp(-p/p_0)$, $p_0=100 $\,TeV. 
 Upper  panel  compares the fluxes calculated with {\tt AAfrag} using  QGSJET-II-04m
 and the ones of \cite{Kelner:2006tc} based on  SIBYLL,
 \cite{Kamae:2006bf}  based on PYTHIA $6.2$, and \cite{Kafexhiu:2014cua}
  based on  GEANT 4.10.0, PYTHIA 8.18, and QGSJET-I. 
Lower panel shows the ratio of  the fluxes for the different parametrizations 
  to the one of   {\tt AAfrag}.}
\label{fig:gamma_E2}
\end{figure}

\begin{figure}
  \includegraphics[width=\linewidth]{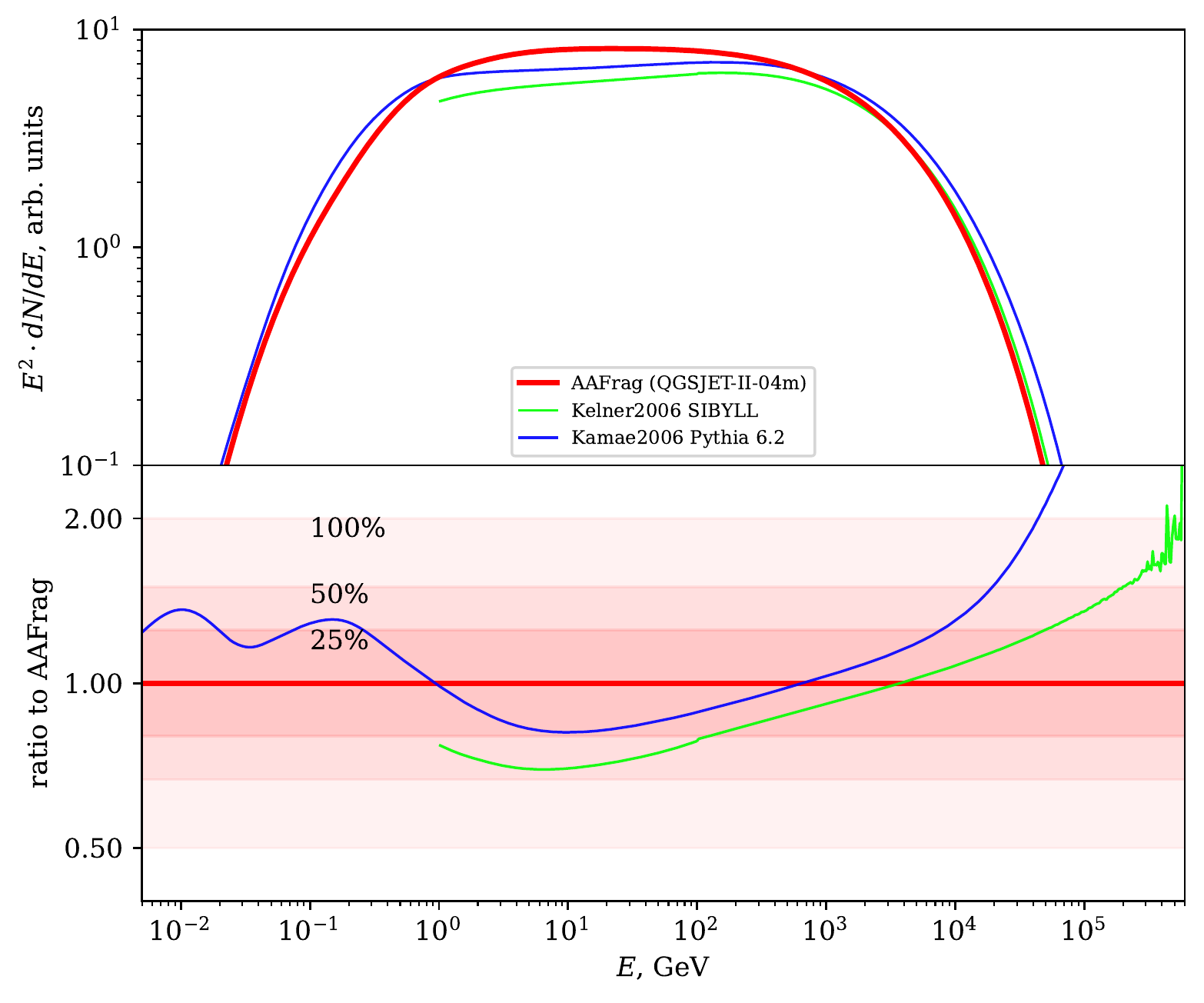}
\caption{All flavor neutrino flux for a power-law cosmic ray spectrum  
($\propto 1/p^2 \exp (-p/p_0)$).}
\label{fig:nu_mu_E2}
\end{figure}

In view of the differences between the predictions of {\tt AAfrag} and 
of the other parametrizations, for monoenergetic protons, we expect a corresponding
spread both for the normalisation and the shapes of the spectra of
gamma-rays and neutrinos also for broad energy distributions of primary
protons. 

In Fig.~\ref{fig:gamma_E2}, we compare the spectra of secondary gamma-rays,
calculated using {\tt AAfrag}, with the those obtained from the
parametrizations.
One can see that the overall normalisation of the predicted gamma-ray fluxes 
varies within $\pm 25\%$ in a wide energy range. In addition, for the 
parametrizations of Kelner~{\it et al.\/} and of Kafexhiu~{\it et al.\/}
based on GEANT~4.10.0, we observe a slightly harder energy slope than for
{\tt AAfrag}. In the particular case of the parametrization of
Kelner~{\it et al.\/}, this is explained by a too steep energy rise of the
inelastic proton-proton cross section, which is at variance with the
respective LHC data. On the other hand, the largest differences between the
various predictions concern the spectral shape in the vicinity of the
high-energy cut-off, where the  spread reaches a factor of two.  This is not
surprising since the steep fall-down of the primary proton 
spectrum in the cutoff region greatly enhances the sensitivity of the results
to very forward ($E_{\gamma} \rightarrow E_p$) photon production in
proton-proton interactions, for which substantial differences between the
model predictions have been observed in the previous section
(c.f.\ Figs.~\ref{fig:gammaTeV} and \ref{fig:gamma100TeV}). 
\change{While the harder gamma-ray fluxes predicted by Kamae~{\it et al.\/}
  is caused by deficiencies in the modeling of the diffractive scattering,
  the harder slope of the pion spectrum predicted by SIBYLL 2.1 results in
  a corresponding  rise of the photon spectrum in the parametrisations of
  Kelner~{\it et al.\/} and  Kafexhiu~{\it et al.\/}.}
This may have  important consequences for the modelling of
the new population of PeV gamma-ray sources recently discovered by 
Tibet-AS$\gamma$~\cite{Amenomori:2021gmk} and LHAASO~\cite{LH}.

In  Fig.~\ref{fig:nu_mu_E2}, we show the comparison of the all-flavor neutrino
spectra  obtained with {\tt AAfrag} and the different parametrizations. Here,
{\tt AAfrag} differs from the other models even stronger. In particular,
{\tt AAfrag} predicts an up to 50\% higher flux of neutrinos, over a broad
energy range except the cutoff region, compared to the model of
Kelner~{\it et al.}  The agreement with Kamae~{\it et al.\/} is better,
with differences within 25\%.  However, the high-energy cut-off in the
neutrino spectra of  Kamae {\it et al.\/} is much sharper. This implies
that the model predictions for neutrino events might differ by up to a
factor of two.

\subsection{Electron and positron spectra}\label{sec:electron_spectra}

\begin{figure}
  \includegraphics[width=\linewidth]{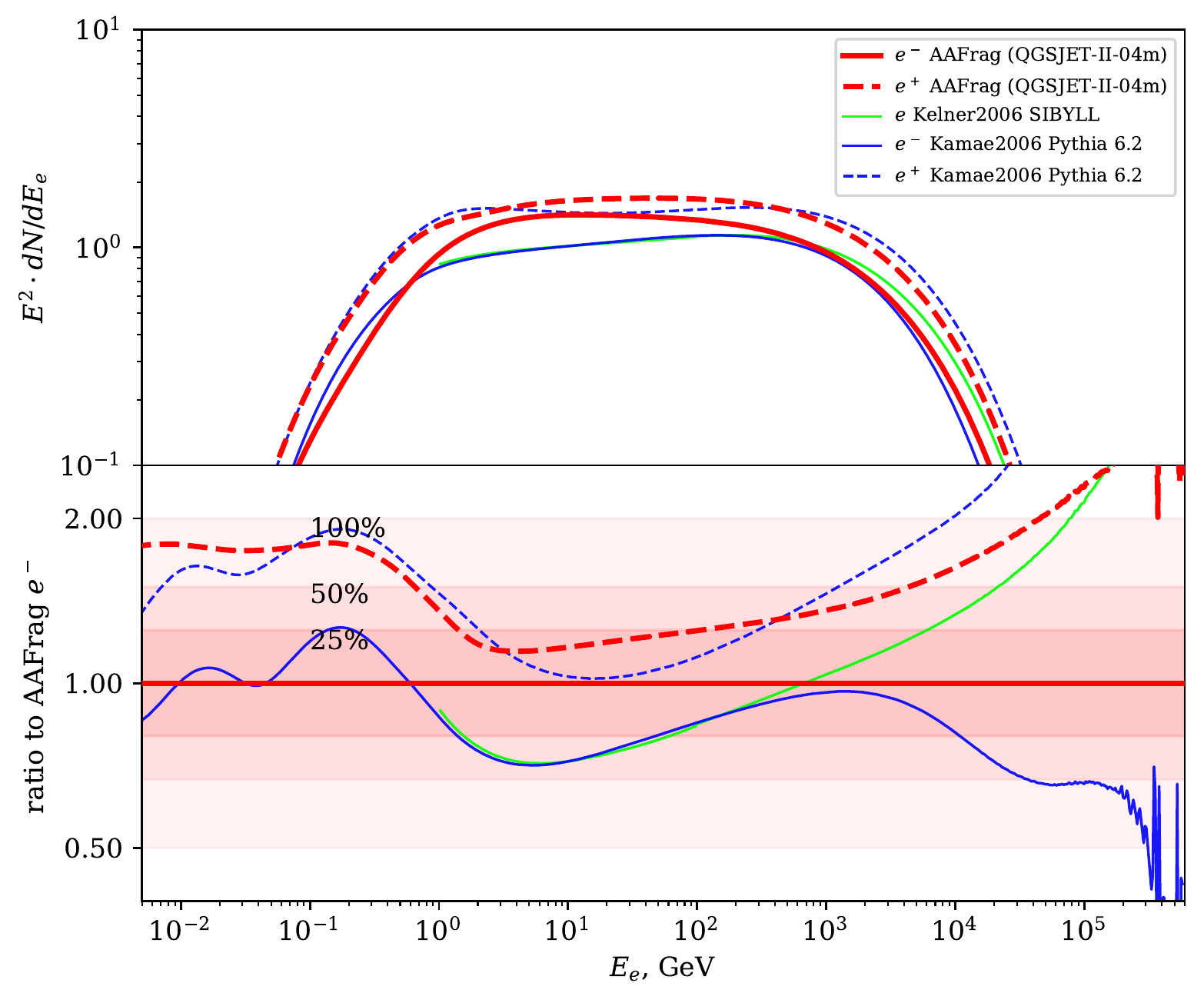}
  \caption{Electron and positron fluxes for a power-law cosmic ray spectrum ($\propto 1/p^2 \exp (-p/p_0)$)
  \label{fig:el_pos_E2}.
  }
\end{figure}

In the upper panel of Fig.~\ref{fig:el_pos_E2},  we show the electron and positron
fluxes  produced by the modelled power-law cosmic-ray spectrum and calculated with  {\tt AAfrag}
and   Kamae~{\it et al.}, together with the average electron-positron
spectrum from Kelner~{\it et al.} The lower panel shows the ratio
of these spectra to the electron spectrum calculated with {\tt AAfrag}.
We note first that both the normalisation and the shape of the positron
and electron spectra differ. While in {\tt AAfrag} there are $\simeq 20\%$
more positrons than electrons at few~GeV, this surplus exceeds a factor
two close to the cutoff. As a result, the positron-electron ratio from
hadronic interactions increases with energy towards the cutoff, an
effect which is missed using the  average electron-positron spectrum
from Kelner~{\it et al.}

\section{aafragpy Python package}
\label{sec:py}

So far we explored differences between the results of the various
parameterisations,
imposing a  sharp low-energy cut-off at the primary particle energy 
$E_p=4$\,GeV, given that the energy range of {\tt AAfrag} is constrained to $E_p>4$\,GeV. Realistic astrophysical source spectra
generally extend to lower energies and {\tt AAfrag} cannot be directly used in such settings. 

To overcome this limitation, we have developed a Python package that implements a Python interface to {\tt AAfrag} and allows to complement {\tt AAfrag} based calculations of particle production spectra with calculations based on parameterisations of low energy production cross-sections. The package can be installed through the standard {\it pip} installer ({\tt pip install aafragpy}). It provides several functions:
\begin{itemize}
\item The matrix of differential cross-sections of secondary particle production in nuclear interaction for a given combination of primary and target nuclei for a given range of energies (function {\tt get\_cross\_section});
\item The differential spectrum of secondary particles produced in the interaction of a given primary spectrum with target nuclei (function {\tt get\_spectrum}).
\end{itemize}

We have included two alternative parameterisations of production cross-sections: those of \citet{Kamae:2006bf} and of \citet{Kafexhiu:2014cua}. The functions for the calculation of differential cross-sections and spectra for these parameterisations are supplemented by {\tt \_Kamae2006} and {\tt \_Kafexhiu2014} suffixes. Both parameterisations can be used only for the calculation of $pp$ cross-sections.  

\begin{figure}
  \includegraphics[width=\linewidth]{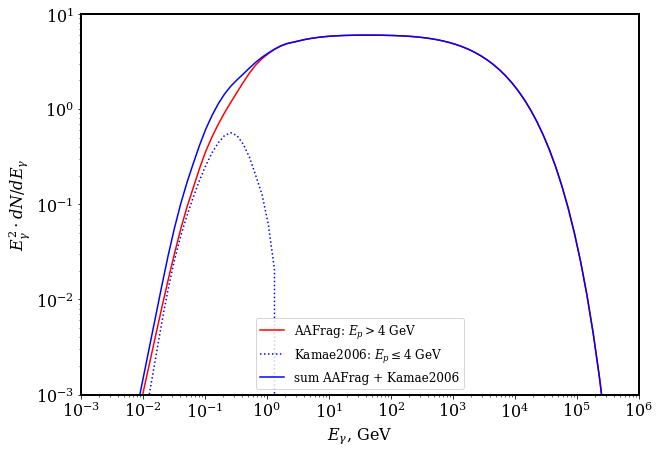}
\caption{Example of the spectrum of $\gamma$-ray emission from $pp$ interactions for an $dN/dp\propto p^{-2}$ powerlaw proton with cutoff on $p_0$=100 TeV. The spectrum generated by {\tt AAfrag} for proton energies $>4$~GeV is complemented by the spectrum calculated using the parameterisations of \citet{Kamae:2006bf} for $E_p<4$\,GeV. An url to interactive Python notebook for the calculation of this spectrum can be found at GitHub page of the project: \url{https://github.com/aafragpy/aafragpy}}
\label{fig:example}
\end{figure}

Figure~\ref{fig:example} shows an example of calculation of the spectrum of $\gamma$-rays from proton-proton interactions, for a power-law proton spectrum with exponential cutoff using such a combined approach. The {\tt AAfrag} production spectrum calculated for protons with energies $E_p>4$\,GeV is shown by the red solid line. The spectrum is complemented by the spectrum of $\gamma$-rays produced by protons with energies $E\le 4$\,GeV, calculated using the \citet{Kamae:2006bf} parameterisation, shown by the blue dotted line. The sum of the two spectra, shown by the blue solid line, corresponds to the total $\gamma$-ray spectrum from a proton distribution in a broad energy range. 

\section{Conclusions}

We have compared the predictions of {\tt AAfrag} for the spectra of
secondaries  produced in proton-proton collisions to those of three
often used parameterisations. We have found considerable differences
both in the normalisation and the shape of the energy spectra,
especially in the region of large energy transfer.
A part of these variations is caused by the (unnecessary) procedure of
fitting the results of QCD based event generators. More importantly,
several parametrisations are based on outdated pre-LHC event generators.
In the case of the energy spectra of photons, we have argued that LHCf
results on the forward production of photons favor the use of the
QGSJET-II-04 model on which {\tt AAfrag} is based on. We have also
stressed that the use of the average electron-positron spectrum
misses the increase with energy of the positron-electron ratio in
hadronic interactions.

The differences found in the normalisation of the secondary spectra
have important implications in the  context of multi-messenger astronomy,
in particular, for the prediction of neutrino fluxes based on gamma-ray
flux measurements, or regarding the inference of the cosmic ray spectrum
based on gamma-ray data.

We also present the easy-to-use Python package {\tt aafragpy}, which allows calculating the differential cross-section and the spectrum of secondary particles produced in nucleus-nucleus interactions in the astrophysical sources. The package uses the calculations of the original {\tt AAfrag} together with calculations performed by Kamae et al. \cite{Kamae:2006bf} and Kafexhiu et al. \cite{Kafexhiu:2014cua} for low-energy secondary production.
\acknowledgments
S.O.\ acknowledges support from the Deutsche Forschungsgemeinschaft 
(project number 465275045).
S.K.\ acknowledges support from the Ministry of Science and Higher Education of the Russian Federation under project ``Fundamental problems of cosmic rays and dark matter" No. 0723-2020-0040.


\begin{thebibliography}{26}
\expandafter\ifx\csname natexlab\endcsname\relax\def\natexlab#1{#1}\fi
\expandafter\ifx\csname bibnamefont\endcsname\relax
  \def\bibnamefont#1{#1}\fi
\expandafter\ifx\csname bibfnamefont\endcsname\relax
  \def\bibfnamefont#1{#1}\fi
\expandafter\ifx\csname citenamefont\endcsname\relax
  \def\citenamefont#1{#1}\fi
\expandafter\ifx\csname url\endcsname\relax
  \def\url#1{\texttt{#1}}\fi
\expandafter\ifx\csname urlprefix\endcsname\relax\def\urlprefix{URL }\fi
\providecommand{\bibinfo}[2]{#2}
\providecommand{\eprint}[2][]{\url{#2}}

\bibitem[{\citenamefont{Spurio}(2018)}]{Spurio:2018knn}
\bibinfo{author}{\bibfnamefont{M.}~\bibnamefont{Spurio}},
  \emph{\bibinfo{title}{{Probes of Multimessenger Astrophysics}}}, Astron.
  Astrophys. Lib. (\bibinfo{publisher}{Springer}, \bibinfo{year}{2018}).

\bibitem[{\citenamefont{Kachelrie{\ss} and
  Semikoz}(2019)}]{Kachelriess:2019oqu}
\bibinfo{author}{\bibfnamefont{M.}~\bibnamefont{Kachelrie{\ss}}}
  \bibnamefont{and} \bibinfo{author}{\bibfnamefont{D.~V.}
  \bibnamefont{Semikoz}}, \bibinfo{journal}{Prog. Part. Nucl. Phys.}
  \textbf{\bibinfo{volume}{109}}, \bibinfo{pages}{103710}
  (\bibinfo{year}{2019}), \eprint{1904.08160}.

\bibitem[{\citenamefont{Aguilar et~al.}(2021)}]{Aguilar:2021tos}
\bibinfo{author}{\bibfnamefont{M.}~\bibnamefont{Aguilar}} \bibnamefont{et~al.}
  (\bibinfo{collaboration}{AMS}), \bibinfo{journal}{Phys. Rept.}
  \textbf{\bibinfo{volume}{894}}, \bibinfo{pages}{1} (\bibinfo{year}{2021}).

\bibitem[{\citenamefont{Bai et~al.}(2019)}]{Bai:2019khm}
\bibinfo{author}{\bibfnamefont{X.}~\bibnamefont{Bai}} \bibnamefont{et~al.}
  (\bibinfo{year}{2019}), \eprint{1905.02773}.

\bibitem[{\citenamefont{Acharya et~al.}(2018)}]{CTAConsortium:2018tzg}
\bibinfo{author}{\bibfnamefont{B.~S.} \bibnamefont{Acharya}}
  \bibnamefont{et~al.} (\bibinfo{collaboration}{CTA Consortium}),
  \emph{\bibinfo{title}{{Science with the Cherenkov Telescope Array}}}
  (\bibinfo{publisher}{WSP}, \bibinfo{year}{2018}), ISBN
  \bibinfo{isbn}{978-981-327-008-4}, \eprint{1709.07997}.

\bibitem[{\citenamefont{Aartsen et~al.}(2021)}]{Aartsen:2020fgd}
\bibinfo{author}{\bibfnamefont{M.~G.} \bibnamefont{Aartsen}}
  \bibnamefont{et~al.} (\bibinfo{collaboration}{IceCube-Gen2}),
  \bibinfo{journal}{J. Phys. G} \textbf{\bibinfo{volume}{48}},
  \bibinfo{pages}{060501} (\bibinfo{year}{2021}), \eprint{2008.04323}.

\bibitem[{\citenamefont{d'Enterria et~al.}(2011)\citenamefont{d'Enterria,
  Engel, Pierog, Ostapchenko, and Werner}}]{dEnterria:2011twh}
\bibinfo{author}{\bibfnamefont{D.}~\bibnamefont{d'Enterria}},
  \bibinfo{author}{\bibfnamefont{R.}~\bibnamefont{Engel}},
  \bibinfo{author}{\bibfnamefont{T.}~\bibnamefont{Pierog}},
  \bibinfo{author}{\bibfnamefont{S.}~\bibnamefont{Ostapchenko}},
  \bibnamefont{and} \bibinfo{author}{\bibfnamefont{K.}~\bibnamefont{Werner}},
  \bibinfo{journal}{Astropart. Phys.} \textbf{\bibinfo{volume}{35}},
  \bibinfo{pages}{98} (\bibinfo{year}{2011}), \eprint{1101.5596}.

\bibitem[{\citenamefont{Ostapchenko}(2011)}]{Ostapchenko:2010vb}
\bibinfo{author}{\bibfnamefont{S.}~\bibnamefont{Ostapchenko}},
  \bibinfo{journal}{Phys. Rev.} \textbf{\bibinfo{volume}{D83}},
  \bibinfo{pages}{014018} (\bibinfo{year}{2011}), \eprint{1010.1869}.

\bibitem[{\citenamefont{Ostapchenko}(2013)}]{Ostapchenko:2013pia}
\bibinfo{author}{\bibfnamefont{S.}~\bibnamefont{Ostapchenko}},
  \bibinfo{journal}{EPJ Web Conf.} \textbf{\bibinfo{volume}{52}},
  \bibinfo{pages}{02001} (\bibinfo{year}{2013}).

\bibitem[{\citenamefont{Kachelrie{\ss}
  et~al.}(2015)\citenamefont{Kachelrie{\ss}, Moskalenko, and
  Ostapchenko}}]{Kachelriess:2015wpa}
\bibinfo{author}{\bibfnamefont{M.}~\bibnamefont{Kachelrie{\ss}}},
  \bibinfo{author}{\bibfnamefont{I.~V.} \bibnamefont{Moskalenko}},
  \bibnamefont{and} \bibinfo{author}{\bibfnamefont{S.~S.}
  \bibnamefont{Ostapchenko}}, \bibinfo{journal}{Astrophys. J.}
  \textbf{\bibinfo{volume}{803}}, \bibinfo{pages}{54} (\bibinfo{year}{2015}),
  \eprint{1502.04158}.

\bibitem[{\citenamefont{Kachelrie\ss{}
  et~al.}(2019)\citenamefont{Kachelrie\ss{}, Moskalenko, and
  Ostapchenko}}]{Kachelriess:2019ifk}
\bibinfo{author}{\bibfnamefont{M.}~\bibnamefont{Kachelrie\ss{}}},
  \bibinfo{author}{\bibfnamefont{I.~V.} \bibnamefont{Moskalenko}},
  \bibnamefont{and}
  \bibinfo{author}{\bibfnamefont{S.}~\bibnamefont{Ostapchenko}},
  \bibinfo{journal}{Comput. Phys. Commun.} \textbf{\bibinfo{volume}{245}},
  \bibinfo{pages}{106846} (\bibinfo{year}{2019}), \eprint{1904.05129}.

\bibitem[{AAf()}]{AAfrag}
\emph{\bibinfo{title}{{AAfrag:} interpolation routines for secondary
  production}}, \bibinfo{note}{available at
  \url{https://sourceforge.net/projects/aafrag/}.}

\bibitem[{\citenamefont{Kamae et~al.}(2006)\citenamefont{Kamae, Karlsson,
  Mizuno, Abe, and Koi}}]{Kamae:2006bf}
\bibinfo{author}{\bibfnamefont{T.}~\bibnamefont{Kamae}},
  \bibinfo{author}{\bibfnamefont{N.}~\bibnamefont{Karlsson}},
  \bibinfo{author}{\bibfnamefont{T.}~\bibnamefont{Mizuno}},
  \bibinfo{author}{\bibfnamefont{T.}~\bibnamefont{Abe}}, \bibnamefont{and}
  \bibinfo{author}{\bibfnamefont{T.}~\bibnamefont{Koi}},
  \bibinfo{journal}{Astrophys. J.} \textbf{\bibinfo{volume}{647}},
  \bibinfo{pages}{692} (\bibinfo{year}{2006}), \bibinfo{note}{[Erratum:
  Astrophys.J. 662, 779 (2007)]}, \eprint{astro-ph/0605581}.

\bibitem[{\citenamefont{Kafexhiu et~al.}(2014)\citenamefont{Kafexhiu,
  Aharonian, Taylor, and Vila}}]{Kafexhiu:2014cua}
\bibinfo{author}{\bibfnamefont{E.}~\bibnamefont{Kafexhiu}},
  \bibinfo{author}{\bibfnamefont{F.}~\bibnamefont{Aharonian}},
  \bibinfo{author}{\bibfnamefont{A.~M.} \bibnamefont{Taylor}},
  \bibnamefont{and} \bibinfo{author}{\bibfnamefont{G.~S.} \bibnamefont{Vila}},
  \bibinfo{journal}{Phys. Rev. D} \textbf{\bibinfo{volume}{90}},
  \bibinfo{pages}{123014} (\bibinfo{year}{2014}), \eprint{1406.7369}.

\bibitem[{\citenamefont{Kelner et~al.}(2006)\citenamefont{Kelner, Aharonian,
  and Bugayov}}]{Kelner:2006tc}
\bibinfo{author}{\bibfnamefont{S.~R.} \bibnamefont{Kelner}},
  \bibinfo{author}{\bibfnamefont{F.~A.} \bibnamefont{Aharonian}},
  \bibnamefont{and} \bibinfo{author}{\bibfnamefont{V.~V.}
  \bibnamefont{Bugayov}}, \bibinfo{journal}{Phys. Rev. D}
  \textbf{\bibinfo{volume}{74}}, \bibinfo{pages}{034018}
  (\bibinfo{year}{2006}), \bibinfo{note}{[Erratum: Phys.Rev.D 79, 039901(E)
  (2009)]}, \eprint{astro-ph/0606058}.

\bibitem[{\citenamefont{Kachelrie{\ss} and
  Ostapchenko}(2012)}]{Kachelriess:2012fz}
\bibinfo{author}{\bibfnamefont{M.}~\bibnamefont{Kachelrie{\ss}}}
  \bibnamefont{and}
  \bibinfo{author}{\bibfnamefont{S.}~\bibnamefont{Ostapchenko}},
  \bibinfo{journal}{Phys. Rev.} \textbf{\bibinfo{volume}{D86}},
  \bibinfo{pages}{043004} (\bibinfo{year}{2012}), \eprint{1206.4705}.

\bibitem[{\citenamefont{Kalmykov and Ostapchenko}(1993)}]{Kalmykov:1993qe}
\bibinfo{author}{\bibfnamefont{N.~N.} \bibnamefont{Kalmykov}} \bibnamefont{and}
  \bibinfo{author}{\bibfnamefont{S.~S.} \bibnamefont{Ostapchenko}},
  \bibinfo{journal}{Phys. Atom. Nucl.} \textbf{\bibinfo{volume}{56}},
  \bibinfo{pages}{346} (\bibinfo{year}{1993}).

\bibitem[{\citenamefont{Kachelrie{\ss}
  et~al.}(2014)\citenamefont{Kachelrie{\ss}, Moskalenko, and
  Ostapchenko}}]{Kachelriess:2014mga}
\bibinfo{author}{\bibfnamefont{M.}~\bibnamefont{Kachelrie{\ss}}},
  \bibinfo{author}{\bibfnamefont{I.~V.} \bibnamefont{Moskalenko}},
  \bibnamefont{and} \bibinfo{author}{\bibfnamefont{S.~S.}
  \bibnamefont{Ostapchenko}}, \bibinfo{journal}{Astrophys. J.}
  \textbf{\bibinfo{volume}{789}}, \bibinfo{pages}{136} (\bibinfo{year}{2014}),
  \eprint{1406.0035}.

\bibitem[{\citenamefont{Adriani et~al.}(2011)}]{Adriani:2011nf}
\bibinfo{author}{\bibfnamefont{O.}~\bibnamefont{Adriani}} \bibnamefont{et~al.}
  (\bibinfo{collaboration}{LHCf}), \bibinfo{journal}{Phys. Lett.}
  \textbf{\bibinfo{volume}{B703}}, \bibinfo{pages}{128} (\bibinfo{year}{2011}),
  \eprint{1104.5294}.

\bibitem[{\citenamefont{Adriani et~al.}(2012)}]{Adriani:2012fz}
\bibinfo{author}{\bibfnamefont{O.}~\bibnamefont{Adriani}} \bibnamefont{et~al.}
  (\bibinfo{collaboration}{LHCf}), \bibinfo{journal}{Phys. Lett.}
  \textbf{\bibinfo{volume}{B715}}, \bibinfo{pages}{298} (\bibinfo{year}{2012}),
  \eprint{1207.7183}.

\bibitem[{\citenamefont{Adriani et~al.}(2016)}]{Adriani:2015iwv}
\bibinfo{author}{\bibfnamefont{O.}~\bibnamefont{Adriani}} \bibnamefont{et~al.}
  (\bibinfo{collaboration}{LHCf}), \bibinfo{journal}{Phys. Rev.}
  \textbf{\bibinfo{volume}{D94}}, \bibinfo{pages}{032007}
  (\bibinfo{year}{2016}), \eprint{1507.08764}.

\bibitem[{\citenamefont{Adriani et~al.}(2018)}]{Adriani:2017jys}
\bibinfo{author}{\bibfnamefont{O.}~\bibnamefont{Adriani}} \bibnamefont{et~al.}
  (\bibinfo{collaboration}{LHCf}), \bibinfo{journal}{Phys. Lett.}
  \textbf{\bibinfo{volume}{B780}}, \bibinfo{pages}{233} (\bibinfo{year}{2018}),
  \eprint{1703.07678}.

\bibitem[{\citenamefont{Stecker}(1971)}]{Stecker71}
\bibinfo{author}{\bibfnamefont{F.}~\bibnamefont{Stecker}},
  \emph{\bibinfo{title}{{Cosmic Gamma Rays}}} (\bibinfo{publisher}{Mono Book
  Corporation, Baltimore}, \bibinfo{year}{1971}).

\bibitem[{\citenamefont{Bell}(2013)}]{Bell:2013vxa}
\bibinfo{author}{\bibfnamefont{A.~R.} \bibnamefont{Bell}},
  \bibinfo{journal}{Astropart. Phys.} \textbf{\bibinfo{volume}{43}},
  \bibinfo{pages}{56} (\bibinfo{year}{2013}).

\bibitem[{\citenamefont{Amenomori et~al.}(2021)}]{Amenomori:2021gmk}
\bibinfo{author}{\bibfnamefont{M.}~\bibnamefont{Amenomori}}
  \bibnamefont{et~al.} (\bibinfo{collaboration}{Tibet ASgamma}),
  \bibinfo{journal}{Phys. Rev. Lett.} \textbf{\bibinfo{volume}{126}},
  \bibinfo{pages}{141101} (\bibinfo{year}{2021}), \eprint{2104.05181}.

\bibitem[{\citenamefont{Cao et~al.}(2021)\citenamefont{Cao, Aharonian, An,
  Axikegu, Bai, Bai, Bao, Bastieri, Bi, Bi et~al.}}]{LH}
\bibinfo{author}{\bibfnamefont{Z.}~\bibnamefont{Cao}},
  \bibinfo{author}{\bibfnamefont{F.~A.} \bibnamefont{Aharonian}},
  \bibinfo{author}{\bibfnamefont{Q.}~\bibnamefont{An}},
  \bibinfo{author}{\bibnamefont{Axikegu}},
  \bibinfo{author}{\bibfnamefont{L.~X.} \bibnamefont{Bai}},
  \bibinfo{author}{\bibfnamefont{Y.~X.} \bibnamefont{Bai}},
  \bibinfo{author}{\bibfnamefont{Y.~W.} \bibnamefont{Bao}},
  \bibinfo{author}{\bibfnamefont{D.}~\bibnamefont{Bastieri}},
  \bibinfo{author}{\bibfnamefont{X.~J.} \bibnamefont{Bi}},
  \bibinfo{author}{\bibfnamefont{Y.~J.} \bibnamefont{Bi}},
  \bibnamefont{et~al.}, \bibinfo{journal}{Nature}
  \textbf{\bibinfo{volume}{594}}, \bibinfo{pages}{33} (\bibinfo{year}{2021}),
  ISSN \bibinfo{issn}{14764687}.

\end{thebibliography}

\end{document}